\documentclass[12pt]{iopart}

\usepackage{cite}

\usepackage{graphicx}

\usepackage{dcolumn}

\begin{document}

\title[Moment-method perturbation theory]{Perturbation theory by the moment method and point-group symmetry}
\author{Francisco M. Fern\'{a}ndez}

\address{INIFTA (UNLP, CCT La Plata-CONICET), Divisi\'on Qu\'imica
Te\'orica, Blvd. 113 S/N,  Sucursal 4, Casilla de Correo 16, 1900
La Plata, Argentina}\ead{fernande@quimica.unlp.edu.ar}

\maketitle

\begin{abstract}
We analyze earlier applications of perturbation theory by the moment method
(also called inner product method) to anharmonic oscillators. For
concreteness we focus on two-dimensional models with symmetry $C_{4v}$ and $%
C_{2v}$ and reveal the reason why some of those earlier treatments proved
unsuitable for the calculation of the perturbation corrections for some
excited states. Point-group symmetry enables one to predict which states
require special treatment.
\end{abstract}

\section{Introduction}

\label{sec:intro}

Many years ago there was great interest in perturbation theory
without wavefunction. Fern\'{a}ndez and Castro\cite{FC84}
developed an approach for multidimensional nonseparable problems
that was based on the recurrence relations of the moments of the
wavefunction instead of the wavefunction itself. They applied it
to the Zeeman effect in Hydrogen and to the Hydrogen molecular
ion. Austin\cite{A84} resorted to this approach in order to obtain
the coefficients of the renormalized perturbation series for the
lowest states of the Zeeman effect in Hydrogen and Arteca et
al\cite{AFMC84} carried out a similar calculation by means of an
order-dependent mapping. Fern\'{a}ndez and Castro\cite{FC85} also
applied this moment perturbation theory (MPT) to the Hydrogen in
parallel magnetic and electric fields and outlined its application
to multidimensional anharmonic oscillators.

Sometime later Killingbeck et al\cite{KJT85} rediscovered the approach and
baptized it inner-product method. They applied it to one-dimensional
anharmonic oscillators and later Killingbeck and Jones\cite{KJ86} to two
dimensional ones. Fern\'{a}ndez et al\cite{FTO87} developed the MPT in a
more general way and showed the general conditions that the recurrence
relations for the moments should satisfy in order to be suitable for a
successful perturbation calculation.

After that, there has been many applications of the MPT to a
variety of quantum-mechanical models\cite
{W91a,W91b,W92a,W92b,W92c,F92,FM92a,FM92b,FO93,W93,WK93,WK94,W95,RDF97,RDF98,W97}
and Fern\'{a}ndez\cite{F01} reviewed some of them in a
comprehensive way. It is worth noting that after so many years of
application of the MPT to such a wide variety of problems
Killingbeck, Witwit and coworkers\cite
{KJT85,KJ86,W91a,W91b,W92a,W92b,W92c,W93,WK93,W95,WK94,W97} never
acknowledged the existence of the seminal articles that first
introduced the MPT\cite{FC84} and its combination with the
renormalized series\cite {A84,AFMC84}.

The simplest two-dimensional model first treated by Killingbeck and Jones%
\cite{KJ86} and Witwit\cite{W91b} is $H=p_{x}^{2}+p_{y}^{2}+x^{2}+y^{2}+%
\lambda \left( ax^{4}+by^{4}+2cx^{2}y^{2}\right) $. Their
calculations were restricted to $a=b$ because ``The potential is
non-separable but shows a high symmetry; this cuts down the amount
of computation required, although the more general anisotropic
case can also be treated by the method''\cite {KJ86} and ``The
inner product method deals with more general parameter values, but
still requires $a=b$ since the equations used exploits this
symmetry to reduce computation''\cite{W91b}. Fern\'{a}ndez and
Ogilvie\cite {FO93} showed that the application of the MPT to the
case $a\neq b$ is not that trivial because the moments of order
zero for some pairs of states coupled by the perturbation satisfy
quadratic, instead of linear, equations (see also
Fern\'andez\cite{F01}). This is the reason why Killingbeck and
Jones\cite{KJ86} and Witwit\cite{W91b} were unable to apply the
perturbation approach to such non-symmetric anharmonic
interactions. In fact, Fern\'{a}ndez and Morales\cite{FM92b} had
previously overcome a similar difficulty in the treatment of the
Zeeman effect in Hydrogen.

Most of the applications of the MPT to anharmonic oscillators took
into account the symmetry of the eigenfunctions (see, for example,
\cite {KJ86,W91b,RDF97,RDF98}). However, such treatments of
symmetry look rather rudimentary when compared with the more
rigorous approach carried out, for example, by Pullen and
Edmonds\cite{PE81a,PE81b}. Those enlightening papers motivated the
application of point-group symmetry (PGS) to several
multidimensional non-Hermitean anharmonic oscillators that led to
most interesting conclusions\cite{FG14a,FG14b,AFG14a,AFG14c}.

The purpose of this paper is to show why the simple symmetry arguments
invoked by Killingbeck and Jones\cite{KJ86} and Witwit\cite{W91b} were
insufficient to solve the Schr\"{o}dinger equation for the multidimensional
anharmonic oscillators by means of the MPT, except for some particular
states. For simplicity we focus on the two-dimensional anharmonic oscillator
shown above but the same ideas apply to all the other models studied so far.

In section~\-\ref{sec:symmetry} we outline the main ideas of PGS\cite
{T64,C90} that we use in section~\ref{sec:AHO_symmetry} to classify the
different cases given by general choices of potential parameters. In section~%
\ref{eq:eigenf_C4v} we analyze the symmetry of the eigenfunctions for the
case $a=b$ and discuss the effect of the symmetry of the anharmonic
interaction on the perturbation corrections to the eigenvalues. We also
outline the effect of symmetry on the behaviour of the moments of the
wavefunction. We calculate the energy eigenvalues for a particular model and
compare our results with those of Killingbeck and Jones\cite{KJ86} and Witwit%
\cite{W91b}. In section~\ref{sec:case_3} we carry out a similar analysis for
the case $a\neq b$. Finally, in section~\ref{sec:conclusions} we summarize
the main results of the paper and generalize the conclusions drawn in the
preceding sections.

\section{Point-group symmetry}

$\label{sec:symmetry}$

In what follows we summarize a few results of group theory that will be
useful throughout this paper.

The set of unitary transformations $U_{i}$, $i=1,2,\ldots ,h$ that leave a
given Hamiltonian operator $H$ invariant $U_{i}HU_{i}^{\dagger }=H$ form a
group with respect to the composition $U_{i}U_{j}$\cite{T64,C90}. The
invariance is obviously equivalent to $[H,U_{i}]=0$. Clearly, if $\psi $ is
an eigenfunction of $H$ with eigenvalue $E$ then $U_{i}\psi $ is also
eigenfunction with the same eigenvalue as follows from $HU_{i}\psi
=U_{i}H\psi =EU_{i}\psi $.

The eigenfunctions of $H$ are bases for the irreducible
representations (irreps) of the point group $G$ of $H$ and can
therefore be classified
according to them\cite{T64,C90}. Group theory gives us projection operators $%
P^{S}$ such that for any arbitrary function $f$ $P^{S}f$ is basis for the
irreducible representation $S$ unless $P^{S}f=0$. The projection operators
are given by
\begin{equation}
P^{S}=\frac{l_{S}}{h}\sum_{i=1}^{h}\chi ^{S}(U_{i})^{*}U_{i},  \label{eq:PS}
\end{equation}
where $l_{S}$ is the dimension of the irreducible representation
$S$, $h$ is the order of the group and $\chi ^{S}(U_{i})$ is the
character of the operation $U_{i}$ for the irrep $S$\cite
{T64,C90}.

It is well known that there is a one-to-one correspondence between
the
unitary operators $U_{i}$ and unitary matrices $\mathbf{M}_{i}$ such that%
\cite{T64}
\begin{equation}
U_{i}f(\mathbf{x})=f(\mathbf{M}_{i}^{\dagger }\mathbf{x}).  \label{eqUif(x)}
\end{equation}

Any projection operator $P$ is self-adjoint $P^{\dagger }=P$ and
idempotent $P^{2}=P$. If $\psi ^{S}$ is an eigenfunction of $H$
and is basis for the irrep $S$ then
\begin{equation}
\left\langle F\right| \left. \psi ^{S}\right\rangle =\left\langle F\right|
\left. P^{S}\psi ^{S}\right\rangle =\left\langle P^{S}F\right| \left. \psi
^{S}\right\rangle .  \label{eq:<F|Psi^S>}
\end{equation}
This brief introduction to group theory will prove sufficient for
all the discussions in the subsequent sections.

\section{Symmetry of the two-dimensional oscillator}

\label{sec:AHO_symmetry}

As indicated in the introduction we will discuss the application of the MPT
to the two-dimensional anharmonic oscillator
\begin{equation}
H=p_{x}^{2}+p_{y}^{2}+x^{2}+y^{2}+\lambda \left(
ax^{4}+by^{4}+2cx^{2}y^{2}\right) ,  \label{eq:H}
\end{equation}
where $p_{q}=-i\frac{d}{dq}$. We have the following cases:

Case 0: $c=0$. This problem is separable in cartesian coordinates and was
chosen by Killingbeck and Jones\cite{KJ86} to test their algorithms. It is
not relevant for present discussion.

Case 1: $a=b=c$. The potential depends only on $r^{2}=x^{2}+y^{2}$ and the
Schr\"{o}dinger equation is therefore separable in spherical coordinates.
This case was also a benchmark for Killingbeck and Jones\cite{KJ86} but it
is of no interest for present purposes.

Case 2: $a=b\neq c$. This case was studied by both Killingbeck and Jones\cite
{KJ86} and Witwit\cite{W91b} by means of the MPT. We will discuss it in the
present paper. A suitable point group is $C_{4v}$ with the following unitary
operations
\begin{eqnarray}
E &:&(x,y)\rightarrow (x,y)  \nonumber \\
C_{4} &:&(x,y)\rightarrow (y,-x)  \nonumber \\
C_{4}^{3} &:&(x,y)\rightarrow (-y,x)  \nonumber \\
C_{2} &:&(x,y)\rightarrow (-x,-y)  \nonumber \\
\sigma _{v1} &:&(x,y)\rightarrow (x,-y)  \nonumber \\
\sigma _{v2} &:&(x,y)\rightarrow (-x,y)  \nonumber \\
\sigma _{d1} &:&(x,y)\rightarrow (y,x)  \nonumber \\
\sigma _{d2} &:&(x,y)\rightarrow (-y,-x)  \label{eq:C4v_oper}
\end{eqnarray}
where $C_{n}$ is a rotation by and angle $2\pi /n$ and $\sigma $ denotes a
reflexion\cite{T64,C90}. The irreps are $A_{1}$, $A_{2}$, $B_{1}$, $B_{2}$
and $E$; the first four ones one-dimensional and the last one
two-dimensional.

Case 3: $a\neq b$. This case was treated by Fern\'{a}ndez and
Ogilvie\cite {FO93} and Radicioni et al\cite{RDF97,RDF98} and was
avoided by Killingbeck and Jones\cite{KJ86} and Witwit\cite{W91b}.
The point group is $C_{2v}$ with operations
\begin{eqnarray}
E &:&(x,y)\rightarrow (x,y)  \nonumber \\
C_{2} &:&(x,y)\rightarrow (-x,-y)  \nonumber \\
\sigma _{v1} &:&(x,y)\rightarrow (x,-y)  \nonumber \\
\sigma _{v2} &:&(x,y)\rightarrow (-x,y)  \label{eq:C2v_oper}
\end{eqnarray}
and only one-dimensional irreps $A_{1}$, $A_{2}$, $B_{1}$ and $B_{2}$. It is
an Abelian group.

\section{Perturbation theory for Case 2}

\label{sec:case_2}

The eigenfunctions and eigenvalues of the unperturbed oscillator $%
H_{0}=H(\lambda =0)$ are
\begin{eqnarray}
H_{0}\varphi _{m,N-m} &=&E_{N}^{(0)}\varphi _{m,N-m},\;m=0,1,\ldots ,N
\nonumber \\
E_{N}^{(0)} &=&2(N+1),\;N=0,1,\ldots ,  \label{eq:eigen_H0}
\end{eqnarray}
where
\begin{equation}
\varphi _{m,n}(x,y)=\phi _{m}(x)\phi _{n}(y),  \label{eq:varphi_mn}
\end{equation}
and $\phi _{m}(q)$ is an eigenfunction of
$H_{HO}=p_{q}^{2}+q^{2}$. The $N$-th unperturbed eigenvalue is
$(N+1)$-fold degenerate and the perturbation
reduces this degeneracy because the symmetry of $H$ is smaller than that of $%
H_{0}$.

The perturbation splits the set of degenerate eigenfunctions of $H_{0}$ with
eigenvalue $E_{2K}^{(0)}$, $K=0,1,\ldots $, into sets of eigenfunctions of $%
H $ of symmetry $A_{1}$, $B_{1}$, $A_{2}$ and $B_{2}$ and those with
eigenvalue $E_{2K+1}^{(0)}$ into sets of eigenfunctions of symmetry $E$. As
a result, the eigenfunctions of $H$ can be written as linear combinations of
the complete set of eigenfunctions of $H_{0}$ in the following way:
\begin{eqnarray}
\psi ^{A_{1}} &=&\sum_{m=0}^{\infty }\sum_{n=m}^{\infty
}c_{mn}^{A_{1}}\varphi _{2m,2n}^{+}  \nonumber \\
\psi ^{B_{1}} &=&\sum_{m=0}^{\infty }\sum_{n=m+1}^{\infty
}c_{mn}^{B_{1}}\varphi _{2m,2n}^{-}  \nonumber \\
\psi ^{A_{2}} &=&\sum_{m=0}^{\infty }\sum_{n=m+1}^{\infty
}c_{mn}^{A_{2}}\varphi _{2m+1,2n+1}^{-}  \nonumber \\
\psi ^{B_{2}} &=&\sum_{m=0}^{\infty }\sum_{n=m}^{\infty
}c_{mn}^{B_{2}}\varphi _{2m+1,2n+1}^{+}  \nonumber \\
\psi _{eo}^{E} &=&\sum_{m=0}^{\infty }\sum_{n=0}^{\infty }c_{mn}^{E}\varphi
_{2m,2n+1}  \nonumber \\
\psi _{oe}^{E} &=&\sum_{m=0}^{\infty }\sum_{n=0}^{\infty }c_{nm}^{E}\varphi
_{2m+1,2n},  \label{eq:eigenf_C4v}
\end{eqnarray}
where
\begin{equation}
\varphi _{m,n}^{+}=\frac{\sqrt{2-\delta _{mn}}}{2}\left( \varphi
_{m,n}+\varphi _{n,m}\right) ,\;\varphi _{m,n}^{-}=\frac{1}{\sqrt{2}}\left(
\varphi _{m,n}-\varphi _{n,m}\right) .  \label{eq:varphi+-}
\end{equation}
The subscripts $o$ and $e$ stand for even and odd, respectively
and refer to the behaviour of $\psi _{eo}^{E}$ and $\psi
_{oe}^{E}$ with respect to the reflection planes $\sigma _{v1}$
and $\sigma _{v2}$. The derivation of these symmetry-adapted basis
functions by means of the projection operators (\ref {eq:PS}) is
straightforward. With respect to the coefficients of the
eigenfunctions of symmetry $E$ note that if $\psi _{eo}^{E}$ is an
eigenfunction of $H$ then $\sigma _{d1}\psi _{eo}^{E}=\psi
_{oe}^{E}$ is also a linearly independent eigenfunction with the
same eigenvalue; therefore, $\{\psi _{eo}^{E},\psi _{oe}^{E}\}$ is
a basis for this irrep.

The MPT is based on recurrence relations for the moments
\begin{equation}
I_{m,n}=\left\langle f_{m,n}\right| \left. \psi \right\rangle
,\;f_{m,n}=x^{m}y^{n}e^{-\alpha (x^{2}+y^{2})},\;m,n=0,1,\ldots ,
\label{eq:Imn}
\end{equation}
where $\psi $ is an eigenfunction of $H$. Therefore, according to (\ref
{eq:<F|Psi^S>}) we have
\begin{equation}
I_{m,n}=\left\langle f_{m,n}\right| \left. \psi ^{S}\right\rangle
=\left\langle P^{S}f_{m,n}\right| \left. \psi ^{S}\right\rangle ,
\end{equation}
for the irrep $S$. Obviously, $I_{m,n}=0$ if $f_{m,n}$ does not have the
proper symmetry. A straightforward calculation shows that
\begin{eqnarray}
I_{2m,2n}^{A_{1}} &=&\frac{1}{2}\left\langle \left(
f_{2m,2n}+f_{2n,2m}\right) \right| \left. \psi ^{A_{1}}\right\rangle
\nonumber \\
I_{2m,2n}^{B_{1}} &=&\frac{1}{2}\left\langle \left(
f_{2m,2n}-f_{2n,2m}\right) \right| \left. \psi ^{B_{1}}\right\rangle
\nonumber \\
I_{2m+1,2n+1}^{A_{2}} &=&\frac{1}{2}\left\langle \left(
f_{2m+1,2n+1}-f_{2n+1,2m+1}\right) \right| \left. \psi ^{A_{2}}\right\rangle
\nonumber \\
I_{2m+1,2n+1}^{B_{2}} &=&\frac{1}{2}\left\langle \left(
f_{2m+1,2n+1}+f_{2n+1,2m+1}\right) \right| \left. \psi ^{B_{2}}\right\rangle
\nonumber \\
I_{2m,2n+1}^{E} &=&\left\langle f_{2m,2n+1}\right| \left. \psi
_{eo}^{E}\right\rangle   \nonumber \\
I_{2m+1,2n}^{E} &=&\left\langle f_{2m+1,2n}\right| \left. \psi
_{oe}^{E}\right\rangle .  \label{eq:Imn_C4v}
\end{eqnarray}
We appreciate that $I_{2m,2n}^{A_{1}}=I_{2n,2m}^{A_{1}}$, $%
I_{2m,2n}^{B_{1}}=-I_{2n,2m}^{B_{1}}$, etc. is the kind of boundary
conditions taken into account in earlier MPT treatments of the anharmonic
oscillators (see, for example, \cite{KJ86,W91b,RDF97,RDF98}).

In order to illustrate the effect of the symmetry of the perturbation on the
corrections of first order to the energy we simply diagonalize the matrix
representation of the perturbation $\mathbf{H}^{\prime }$ in the basis set
of degenerate eigenfunctions of $H_{0}$. Obviously, since functions of
different symmetry do not mix it is preferable (and advisable) to
diagonalize the matrices $\mathbf{H}^{\prime }(S)$ for every one of the
irreps $S$.

The case $N=0$ is trivial but we include it here for completeness. There is
only one function and the correction of first order is
\begin{equation}
\left\langle \varphi _{00}\right| H^{\prime }\left| \varphi
_{00}\right\rangle =\frac{3a+c}{2}.
\end{equation}

When $N=1$ we have two functions of symmetry $E$ that lead to a diagonal
matrix:
\begin{equation}
\mathbf{H}^{\prime }(E)=\left(
\begin{array}{cc}
\frac{3\left( 3a+c\right) }{2} & 0 \\
0 & \frac{3\left( 3a+c\right) }{2}
\end{array}
\right) ,
\end{equation}
because $\varphi _{0,1}$ and $\varphi _{1,0}$ have different eigenvalues
with respect to the operators $\sigma _{v1}$ and $\sigma _{v2}$.

When $N=2$ the function $\varphi _{11}$ is $B_{2}$ and the functions $%
\{\varphi _{20},\varphi _{02}\}$ are linear combinations of functions of
symmetry $A_{1}$ and $B_{1}$. We have
\begin{equation}
\left\langle \varphi _{11}\right| H^{\prime }\left| \varphi
_{11}\right\rangle =\frac{3\left( 5a+3c\right) }{2},
\end{equation}
for the former and
\begin{equation}
\mathbf{H}^{\prime }=\left(
\begin{array}{cc}
\frac{21a+5c}{2} & c \\
c & \frac{21a+5c}{2}
\end{array}
\right) ,
\end{equation}
for the latter. This matrix is not diagonal; however if we use the
symmetry-adapted functions $\{\varphi _{20}^{+},\varphi
_{20}^{-}\}$ we obtain a diagonal one:
\begin{equation}
\mathbf{H}^{\prime }=\left(
\begin{array}{cc}
\frac{7\left( 3a+c\right) }{2} & 0 \\
0 & \frac{3\left( 7a+c\right) }{2}
\end{array}
\right)
\end{equation}

A different situation arises for $N=3$ because the four degenerate
unperturbed eigenfunctions can be grouped into two pairs of symmetry $E$: $%
\{\varphi _{3,0},\varphi _{1,2}\}$ and $\{\varphi _{0,3},\varphi _{2,1}\}$.
Both lead to identical matrix representations of the perturbation:
\begin{equation}
\mathbf{H}^{\prime }(E)=\left(
\begin{array}{cc}
\frac{39a+7c}{2} & \sqrt{3}c \\
\sqrt{3}c & \frac{3\left( 9a+5c\right) }{2}
\end{array}
\right)
\end{equation}
with eigenvalues

\begin{equation}
E^{(1)}=\frac{11\left( 3a+c\right) \pm
2\sqrt{9a^{2}-12ac+7c^{2}}}{2}.
\end{equation}

We cannot make this matrix diagonal by means of symmetry operations;
therefore the inner product method as applied by Killingbeck and Jones\cite
{KJ86} and Witwit\cite{W91b} is expected to fail, and in fact they entirely
omitted the treatment of these energy levels. By a judicious manipulation of
the moment recurrence relations one can derive a quadratic equation for one
of the moments of order zero in order to obtain the corrections to the energy%
\cite{FO93,RDF97,RDF98,F01}.

From the results above we conclude that the first energy levels corrected to
first order are given by
\begin{eqnarray}
E_{1A_{1}} &=&2+\frac{3a+c}{2}\lambda +\ldots  \nonumber \\
E_{1E} &=&4+\frac{3\left( 3a+c\right) }{2}\lambda +\ldots  \nonumber \\
E_{1B_{2}} &=&6+\frac{3(5a+3c)}{2}\lambda +\ldots  \nonumber \\
E_{2A_{1}} &=&6+\frac{7\left( 3a+c\right) }{2}\lambda +\ldots  \nonumber \\
E_{1B_{1}} &=&6+\frac{3\left( 7a+c\right) }{2}\lambda +\ldots  \nonumber \\
E_{2E} &=&8+\frac{11\left( 3a+c\right) -2\sqrt{9a^{2}-12ac+7c^{2}}}{2}%
\lambda +\ldots  \nonumber \\
E_{3E} &=&8+\frac{11\left( 3a+c\right) +2\sqrt{9a^{2}-12ac+7c^{2}}}{2}%
\lambda +\ldots  \label{eq:E_C4v}
\end{eqnarray}
where the subscript $jS$ indicates that the symmetry $S$ appears for the
j-th time. The level order may change with the relative magnitudes of the
potential parameters.

Killingbeck and Jones\cite{KJ86} and Witwit\cite{W91b} calculated
some of the eigenvalues corresponding to $N=0,1,2,4$,
respectively. Note that they omitted the four states that stem
from $N=3$ and three of the 5 states that come from $N=4$ (which
do not appear in equation (\ref{eq:E_C4v})). The reason is that
their mathematical treatment of the moment recurrence relations
did not enable them to obtain suitable working equations for such
cases. They chose examples where the perturbation corrections of
first order are linear functions of the potential parameters,
while for $N=3$, for example, they are nonlinear as shown above.
In the case $N=3$ one expects quadratic equations for the moments
of order zero as shown by Fern\'{a}ndez
and Ogilvie\cite{FO93} and Radicioni et al\cite{RDF97,RDF98} (see also Fern%
\'{a}ndez\cite{F01}) for the Case 3 discussed in the following section.

\begin{figure}[]
\begin{center}
\includegraphics[width=12cm]{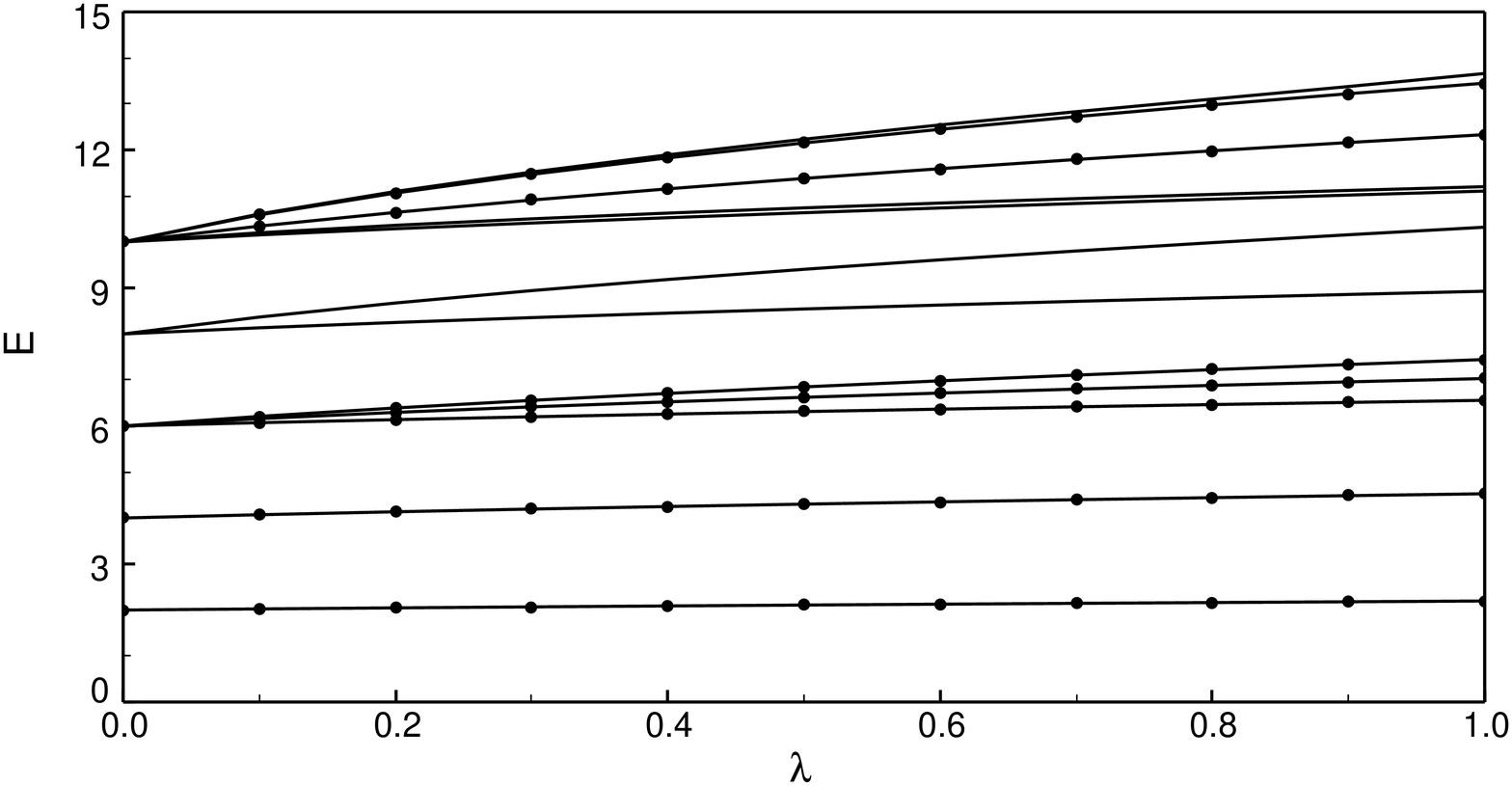}
\par
\end{center}
\caption{First eigenvalues of the anharmonic oscillator
(\ref{eq:H}). Solid lines and points indicate present and Witwit's
results\cite{W91b}, respectively } \label{fig:Case_2}
\end{figure}

In order to illustrate the omission of states mentioned above we
calculated the eigenvalues of the anharmonic oscillator with
$a=b=0$, $c=1$ and $0\leq \lambda \leq 1$ and compared them with
those reported by Witwit\cite{W91b} (Killingbeck and
Jones\cite{KJ86} also showed some results). Figure~\ref
{fig:Case_2} clearly shows that those authors omitted 7 of the 15
states that come from $N=0,1,2,3,4$. In order to obtain the
eigenvalues displayed in that figure we resorted to the
Rayleigh-Ritz variational method in the Krylov space spanned by
the set of non-orthogonal functions $\Omega _{n}^{S}=H^{n}\Omega
^{S}$ for each irrep $S$. For the present problem suitable
reference functions $\Omega ^{S}$ are given by
\begin{eqnarray}
\Omega ^{A_{1}} &=&e^{-\alpha (x^{2}+y^{2})}  \nonumber \\
\Omega ^{B_{1}} &=&(x^{2}-y^{2})e^{-\alpha (x^{2}+y^{2})}  \nonumber \\
\Omega ^{A_{2}} &=&(xy^{3}-x^{3}y)e^{-\alpha (x^{2}+y^{2})}  \nonumber \\
\Omega ^{B_{2}} &=&xye^{-\alpha (x^{2}+y^{2})}  \nonumber \\
\Omega _{oe}^{E} &=&xe^{-\alpha (x^{2}+y^{2})}  \nonumber \\
\Omega _{eo}^{E} &=&ye^{-\alpha (x^{2}+y^{2})},
\end{eqnarray}
where $\alpha >0$ can be chosen in order to get the greatest rate of
convergence. Here we simply chose $\alpha =1$ and matrices of dimension $%
20\times 20$ because great accuracy is not required.

In order to facilitate the comparison of the results and the
interpretation of figure~\ref{fig:Case_2} in what follows we list
the labels used in the present paper and in the ones of those
authors:
\begin{eqnarray}
N &=&0\rightarrow 1A_{1}\rightarrow (0,0,e)  \nonumber \\
N &=&1\rightarrow 1E\rightarrow (0,1,\mathrm{mixed}),\,(1,0,\mathrm{mixed})
\nonumber \\
N &=&2\rightarrow \left\{
\begin{array}{c}
1B_{1}\rightarrow (0,2,\mathrm{o}) \\
2A_{1}\rightarrow (0,2,\mathrm{e}) \\
1B_{2}\rightarrow (1,1,\mathrm{e})
\end{array}
\right.   \nonumber \\
N &=&3\rightarrow \left\{
\begin{array}{c}
2E\rightarrow \mathrm{omitted} \\
3E\rightarrow \mathrm{omitted}
\end{array}
\right.   \nonumber \\
N &=&4\rightarrow \left\{
\begin{array}{c}
3A_{1}\rightarrow \mathrm{omitted} \\
2B_{1}\rightarrow \mathrm{omitted} \\
1A_{2}\rightarrow (1,3,\mathrm{o}) \\
2B_{2}\rightarrow (1,3,\mathrm{e}) \\
4A_{1}\rightarrow \mathrm{omitted}
\end{array}
\right. .
\end{eqnarray}
\newline
Killingbeck and Jones\cite{KJ86} mentioned problems with some excited
states. They briefly referred to the coupling of the states $\varphi _{1,2}$
and $\varphi _{3,0}$ but did not show any equation that could overcome the
difficulty. On the other hand, Witwit\cite{W91b} never considered this
situation at all.

\section{Perturbation theory for Case 3}

\label{sec:case_3}

When $a\neq b$ the perturbation splits the set of degenerate
eigenfunctions of $H_{0}$ with eigenvalue $E_{2K}^{(0)}$,
$K=0,1,\ldots $, into sets of eigenfunctions of $H$ of symmetry
$A_{1}$, and $A_{2}$ and those with eigenvalue $E_{2K+1}^{(0)}$
into sets of eigenfunctions of symmetry $B_{1}$ and $B_{2}$. The
eigenfunctions of $H$ can be written as linear combinations of the
unperturbed eigenfunctions in the following way:
\begin{eqnarray}
\psi ^{A_{1}} &=&\sum_{m=0}^{\infty }\sum_{n=0}^{\infty
}c_{mn}^{A_{1}}\varphi _{2m,2n}  \nonumber \\
\psi ^{A_{2}} &=&\sum_{m=0}^{\infty }\sum_{n=0}^{\infty
}c_{mn}^{A_{2}}\varphi _{2m+1,2n+1}  \nonumber \\
\psi ^{B_{1}} &=&\sum_{m=0}^{\infty }\sum_{n=0}^{\infty
}c_{mn}^{B_{1}}\varphi _{2m+1,2n}  \nonumber \\
\psi ^{B_{2}} &=&\sum_{m=0}^{\infty }\sum_{n=0}^{\infty
}c_{mn}^{B_{2}}\varphi _{2m,2n+1}.  \label{eq:eigenf_C2v}
\end{eqnarray}
As in the preceding case we apply straightforward perturbation theory of
first order beginning with the trivial case $N=0$:
\begin{equation}
\left\langle \varphi _{00}\right| H^{\prime }\left| \varphi
_{00}\right\rangle =\frac{3a+3b+2c}{4}.
\end{equation}

When $N=1$ the two functions exhibit symmetry $B_{1}$ and $B_{2}$ and the
perturbation matrix is diagonal
\begin{equation}
\mathbf{H}^{\prime }=\left(
\begin{array}{cc}
\frac{3\left( 5a+b+2c\right) }{4} & 0 \\
0 & \frac{3\left( a+5b+2c\right) }{4}
\end{array}
\right) .
\end{equation}

Of the three functions for $N=2$ two exhibit symmetry $A_{1}$ and the
remaining one symmetry $A_{2}$. The matrix for the former is not diagonal
\begin{equation}
\mathbf{H}^{\prime }(A_{1})=\left(
\begin{array}{cc}
\frac{3a+39b+10c}{4} & c \\
c & \frac{39a+3b+10c}{4}
\end{array}
\right) ,
\end{equation}
and has eigenvalues
\begin{equation}
E^{(1)}=\frac{21a+21b+10c\pm 2\sqrt{81a^{2}-162ab+81b^{2}+4c^{2}}}{4}.
\end{equation}
This matrix cannot be brought into a diagonal form by means of symmetry
operations. On the other hand, for the symmetry $A_{2}$ we simply have
\begin{equation}
\left\langle \varphi _{11}\right| H^{\prime }\left| \varphi
_{11}\right\rangle =\frac{3\left( 5a+5b+6c\right) }{4}
\end{equation}

Thus, the first eigenvalues corrected to first order are:
\begin{eqnarray}
E_{1A_{1}} &=&2+\frac{3a+3b+2c}{4}\lambda +\ldots   \nonumber \\
E_{1B_{1}} &=&4+\frac{3\left( 5a+b+2c\right) }{4}\lambda +\ldots   \nonumber
\\
E_{1B_{2}} &=&4+\frac{3\left( a+5b+2c\right) }{4}\lambda +\ldots   \nonumber
\\
E_{2A_{1}} &=&6+\frac{21a+21b+10c-2\sqrt{81a^{2}-162ab+81b^{2}+4c^{2}}}{4}%
\lambda +\ldots   \nonumber \\
E_{3A_{1}} &=&6+\frac{21a+21b+10c-2\sqrt{81a^{2}-162ab+81b^{2}+4c^{2}}}{4}%
\lambda +\ldots   \nonumber \\
E_{1A_{2}} &=&6+\frac{3\left( 5a+5b+6c\right) }{4}\lambda +\ldots
\end{eqnarray}
In this case the inner product method as applied by Killingbeck and Jones%
\cite{KJ86} and Witwit\cite{W91b} begins to be unsuitable at $N=2$
because it is not expected to yield the states $A_{1}$ (although,
it is known to be successful for the remaining state $A_{2}$
\cite{F01}). In order to obtain the perturbation corrections for
$E_{2A_{1}}$ and $E_{3A_{1}}$ one has to manipulate the moment
recurrence relations and derive a quadratic expression for one of
the moments of order zero as shown by Fern\'{a}ndez
and Ogilvie\cite{FO93} and Radicioni et al\cite{RDF97,RDF98} (see also Fern%
\'{a}ndez\cite{F01}).

\section{Conclusions}

\label{sec:conclusions}

The aim of this paper is the discussion of the application of the MPT (or
inner product method) to anharmonic oscillators. For concreteness in the
preceding sections we focused on the simple two-dimensional model (\ref{eq:H}%
) but other cases can be treated in the same way. We have shown that PGS is
extremely useful for understanding the way in which the perturbation affects
the states of the unperturbed model and the boundary and initial conditions
that one should consider during the manipulation of the moment recurrence
relations. In particular we were interested in the determination of the
conditions under which the application of the inner product method in the
way proposed by Killingbeck and Jones\cite{KJ86} and Witwit\cite{W91b} is
successful. The conclusion of our analysis is that such an approach is
expected to fail for some excited states of the anharmonic oscillator (\ref
{eq:H}) for the Cases 2 and 3 discussed in sections \ref{sec:case_2} and \ref
{sec:case_3}, respectively. The particular results derived there can be
generalized in the following way: suppose that $\chi _{n,j}$, $j=1,2,\ldots
,\nu $ are degenerate eigenfunctions of $H_{0}$ with eigenvalue $E_{n}^{(0)}$
adapted to the symmetry $S$ of the point group $G$ of $H$. If the dimension $%
l_{S}$ of $S$ is smaller than $\nu $ then the application of the
inner product approach proposed by Killingbeck and
Jones\cite{KJ86} and Witwit\cite {W91b} is expected to fail. In
this case one has to manipulate the recurrence relations for the
moments in order to derive a polynomial function of one of the
moments of order zero. Each of the roots of this polynomial will
lead to the correction to the energy of each of the states coupled
by the perturbation. For example, the states stemming from $N=3$
in Case 2 lead to two quadratic equations from which we obtain the
perturbation corrections for the energy levels $E_{2E}$ and
$E_{3E}$. It is not difficult to verify that more complex
situations appear for greater values of $N$. The Case 3 was not
treated by Killingbeck and Jones\cite{KJ86} and Witwit\cite {W91b}
but Fern\'{a}ndez and Ogilvie\cite{FO93} and Radicioni et al\cite
{RDF97,RDF98} (see also Fern\'{a}ndez\cite{F01}) showed how to
obtain quadratic polynomial equations for the two states $A_{1}$
coming from $N=2$. PGS predicts that for $N=3$ we should have two
quadratic equations that yield the perturbation corrections for
two states of symmetry $B_{1}$ and two states of symmetry $B_{2}$.

\end{document}